\documentstyle[graphicx]{./mn}

\newif\ifAMStwofonts
\def\mnras{MNRAS}
\def\apj{ApJ}
\def\apjl{ApJL}
\def\apjs{ApJS}
\def\aap{A\&A}

\def\aj{AJ}
\def\nat{Nat}

\ifoldfss
  \ifCUPmtlplainloaded \else
    \NewTextAlphabet{textbfit} {cmbxti10} {}
    \NewTextAlphabet{textbfss} {cmssbx10} {}
    \NewMathAlphabet{mathbfit} {cmbxti10} {} % for math mode
    \NewMathAlphabet{mathbfss} {cmssbx10} {} %  "   "    "
  \fi
  \ifAMStwofonts
    \ifCUPmtlplainloaded \else
      \NewSymbolFont{upmath} {eurm10}
      \NewSymbolFont{AMSa} {msam10}
      \NewMathSymbol{\upi}     {0}{upmath}{19}
      \NewMathSymbol{\umu}     {0}{upmath}{16}
      \NewMathSymbol{\upartial}{0}{upmath}{40}
      \NewMathSymbol{\leqslant}{3}{AMSa}{36}
      \NewMathSymbol{\geqslant}{3}{AMSa}{3E}

      \let\leq=\leqslant 
       
    \fi
    \ifCUPmtlplainloaded \else
      \NewSymbolFont{upmath} {eurm10}
      \NewSymbolFont{AMSa} {msam10}
      \NewMathSymbol{\upi}     {0}{upmath}{19}
      \NewMathSymbol{\umu}     {0}{upmath}{16}
      \NewMathSymbol{\upartial}{0}{upmath}{40}
      \NewMathSymbol{\leqslant}{3}{AMSa}{36}
      \NewMathSymbol{\geqslant}{3}{AMSa}{3E}

      \let\leq=\leqslant 
       
    \fi
  \fi
\fi % End of OFSS

\ifnfssone
  \newmathalphabet{\mathit}
  \addtoversion{normal}{\mathit}{cmr}{m}{it}
  \addtoversion{bold}{\mathit}{cmr}{bx}{it}
  \newmathalphabet{\mathbfit} % math mode version of \textbfit{..}
  \addtoversion{normal}{\mathbfit}{cmr}{bx}{it}
  \addtoversion{bold}{\mathbfit}{cmr}{bx}{it}
  \newmathalphabet{\mathbfss} % math mode version of \textbfss{..}
  \addtoversion{normal}{\mathbfss}{cmss}{bx}{n}
  \addtoversion{bold}{\mathbfss}{cmss}{bx}{n}
  \ifAMStwofonts
    \ifCUPmtlplainloaded \else
      \UseAMStwoboldmath
      \makeatletter
      \new@mathgroup\upmath@group
      \define@mathgroup\mv@normal\upmath@group{eur}{m}{n}
      \define@mathgroup\mv@bold\upmath@group{eur}{b}{n}
      \edef\UPM{\hexnumber\upmath@group}
      \new@mathgroup\amsa@group
      \define@mathgroup\mv@normal\amsa@group{msa}{m}{n}
      \define@mathgroup\mv@bold\amsa@group{msa}{m}{n}
      \edef\AMSa{\hexnumber\amsa@group}
      \makeatother
      \mathchardef\upi="0\UPM19
      \mathchardef\umu="0\UPM16
      \mathchardef\upartial="0\UPM40
      \mathchardef\leqslant="3\AMSa36
      \mathchardef\geqslant="3\AMSa3E

      \let\leq=\leqslant 

    \fi
  \fi
\fi % End of NFSS release 1

\ifnfsstwo
  \DeclareMathAlphabet{\mathbfit}{OT1}{cmr}{bx}{it}
  \SetMathAlphabet\mathbfit{bold}{OT1}{cmr}{bx}{it}
  \DeclareMathAlphabet{\mathbfss}{OT1}{cmss}{bx}{n}
  \SetMathAlphabet\mathbfss{bold}{OT1}{cmss}{bx}{n}
  \ifAMStwofonts
    \ifCUPmtlplainloaded \else
      \DeclareSymbolFont{UPM}{U}{eur}{m}{n}
      \SetSymbolFont{UPM}{bold}{U}{eur}{b}{n}
      \DeclareSymbolFont{AMSa}{U}{msa}{m}{n}
      \DeclareMathSymbol{\upi}{0}{UPM}{"19}
      \DeclareMathSymbol{\umu}{0}{UPM}{"16}
      \DeclareMathSymbol{\upartial}{0}{UPM}{"40}
      \DeclareMathSymbol{\leqslant}{3}{AMSa}{"36}
      \DeclareMathSymbol{\geqslant}{3}{AMSa}{"3E}

      \let\leq=\leqslant 

    \fi
  \fi
\fi % End of NFSS release 2

\ifCUPmtlplainloaded \else
  \ifAMStwofonts \else % If no AMS fonts
    \def\upi{\pi}
    \def\umu{\mu}
    \def\upartial{\partial}
  \fi
\fi

\title{The prevalence of FR\,I radio quasars}
\author[Heywood et al.]
{Ian~Heywood, Katherine~M.~Blundell and Steve~Rawlings\\University of Oxford
Astrophysics, Keble Road, Oxford, OX1 3RH, UK}
\date{Accepted 2007 month 00. Received 2007 month 00}
\pagerange{\pageref{firstpage}--\pageref{lastpage}}
\pubyear{200x}

\def\LaTeX{L\kern-.36em\raise.3ex\hbox{a}\kern-.15em
    T\kern-.1667em\lower.7ex\hbox{E}\kern-.125emX}

\begin{document}

\label{firstpage}

\maketitle

\begin{abstract}

We present deep, multi-VLA-configuration radio images for a set of 18 quasars, having redshifts between 0.36 and 2.5,
from the 7C quasar survey. 
Approximately one quarter of these quasars have FR\,I-type twin-jet structures and the
remainder are a broad range of wide angle tail, fat double, classical double, core-jet and hybrid sources.  These
images demonstrate that FR\,I quasars are prevalent in the
universe, rather than non-existent as had been suggested in the
literature prior to the serendipitous discovery of the first FR\,I
quasar a few years ago, the optically powerful `radio quiet'
quasar E\,1821+643.

Some of the FR\,I quasars have radio luminosities exceeding the traditional FR\,I / FR\,II
break luminosity, however we find no evidence for FR\,II quasars with luminosities significantly below the break. We consider whether the existence of such high
luminosity FR\,I structures is due to the increasingly inhomogeneous environments
in the higher redshift universe. 

\end{abstract}


\section{Introduction}

The morphological classification scheme for extended extragalactic radio
sources described by Fanaroff and Riley (1974) divides these objects
into two groups: radio structures decreasing in brightness as the
distance from the core increases are classed as FR\,I 
and sources with bright hotspots at the extremities are
labelled FR\,II. The two types of object have hitherto been
observed to be fairly sharply divided by a luminosity value known as the
FR\,I / FR\,II break, with FR\,IIs being the more luminous. 

Jets from active galactic nuclei generally give rise to
structures which fall into both of these categories although the
literature contains assertions that optically powerful quasars
with FR\,I type radio structures do not exist (Falcke et al.,
1995; Baum et al., 1995). The general view was that quasars either
had luminous FR\,II structures with hotspots or one sided core-jet structure (radio-loud quasars),
or they did not have any significant radio emission other than unresolved cores
(radio-quiet quasars). However, these claims were based purely on
the absence of FR\,I quasars from the bright 3C survey together
with their absence from limited snapshot (i.e.\ few minute) radio images in the
case of optically selected quasars.  The inadequecy and
insufficiency of short snapshot images only sensitive to small
spatial scales were critiqued by Blundell (2005).  

In general, radio observations of quasars to date have consisted
of a short `snapshot' observation using the VLA in the extended
A- or B-array configurations (e.g.\ Miller et al., 1990; Miller et
al, 1993; Kellermann et al., 1994; Hooper et al., 1996; Kukula et
al., 1998; Goldschmidt et al., 1999; White et al.,
2000). Observations of this type would be inadequate for
detecting FR\,I structures for two reasons. First, as the name
would imply the integration time of a snapshot observation is
generally short ($\leq~5$ minutes) which compromises the
sensitivity and ability to perform high-fidelity deconvolution.
Second, the extended configurations of the VLA will resolve out
the extended structures with low intensity gradients, features
which are characteristic of FR\,I sources.  Thus it is at least
plausible that observations to date have not detected smooth,
faint 100-kpc-scale FR\,I structures with quasars simply because
the observations have been inadequate.  

It is also noteworthy that the luminosity of every previously
identified FR\,I source is below or very close to the
FR\,I/FR\,II break luminosity, which at 178~MHz is
$\sim$10$^{25.5}$~W~Hz$^{-1}$~sr$^{-1}$ (Fanaroff \& Riley,
1974). This explains the lack of FR\,I structures associated with
quasars found so far as the FR\,I/FR\,II break luminosity
falls below the flux limit at the
rather low redshift of 0.4, for some of the current deepest low-frequency selected complete samples (Rawlings et
al, 1998; Blundell, Rawlings \& Willott, 1999).

This leads on to the final consideration for the apparent lack of
FR\,I quasars. The rarity of quasars implies that large volumes
of space must be searched in order to observe one or more FR I sources. Generally quasars are found at redshifts greater
than 0.4, and as more distant quasars are observed the
feasibility of detecting a low-luminosity FR I source decreases for a radio flux-limited survey.

Given that jets of both FR\,I and FR\,II morphologies are
common for many types of active galaxy, coupled with
the above observational considerations, it is plausible that the
lack of FR\,I-type quasars is not due to the fact that they do not
exist but because the existing observations and radio surveys are
simply not detecting them. In fact the first \emph{deep} image of
the so-called radio quiet, optically powerful quasar E1821+643
serendipitously revealed a 300~kpc FR\,I jet structure (Blundell
\& Rawlings, 2001).

With this discovery and the limitations of previous observations
firmly in mind VLA observations of 18 quasars selected from the
7C quasar survey (Riley et al., 1999) were scheduled. The objects
selected were FR\,I candidates from the 7C quasar survey
and details of the selection criteria and the observations can be
found in Section\,\ref{sec:observations}. In
Section\,\ref{sec:results} we present combined multi-VLA-configuration radio
images and flux density measurements of the 18-object sample. Section\,\ref{sec:discussion} contains 
morphological discussions on a per-source basis, as well as redshift-dependent considerations both for our
sample and for FR I quasars in general. Concluding remarks are presented in Section\,\ref{sec:conclusions}.

Throughout this paper the $H_{0}$~=~70~km~s$^{-1}$~Mpc$^{-1}$,
$\Omega_{\rm M}$~=~0.3 and $\Omega_{\Lambda}$~=~0.7 cosmology is
used.

\section{Sample Selection, Observations and Data Reduction}
\label{sec:observations}

The starting point for the sample selection was the 7C quasar
survey (Riley et al., 1999). This `filtered sample' of
quasars was chosen from a complete sample selected to be brighter
than 100\,mJy at the low radio frequency of 151\,MHz, and optical colour filters were applied to favour the detection of quasars.

\begin{figure}
\nonumber
\centering
\includegraphics[width= \columnwidth]{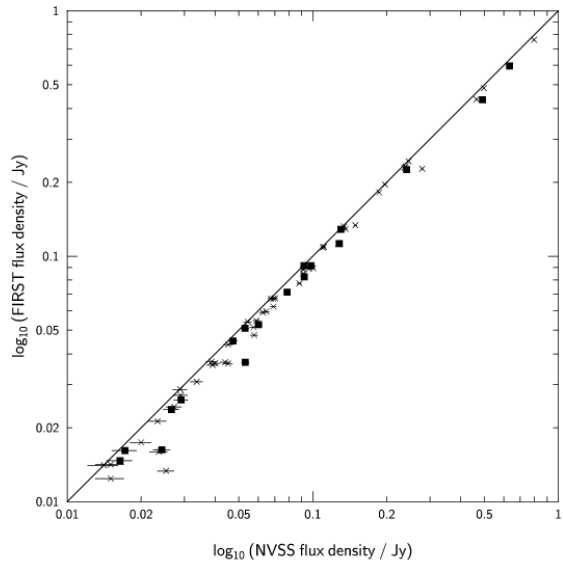}
\caption{\label{fig:NVSS_FIRST}The FIRST against NVSS radio flux density measurements for the confirmed radio quasars from the 7C quasar sample (Riley et al., 1999). Objects selected
for our radio observations are represented by the filled square symbols. Errors for both the FIRST and NVSS flux densities are plotted. The NVSS errors are generally too small to be seen
in the upper half of the plot. Errors in the FIRST flux density measurements are invisible throughout, although it is noteworthy that these are only the quoted RMS values and do not include 
``systematic'' errors (e.g. from missing flux).} 
\end{figure}

Beginning with the Riley et al. sample, we took the bona fide quasars (confirmed via unpublished optical spectra; Rawlings, priv. comm.) and plotted their radio flux 
density measurements from the FIRST survey (Becker et al., 1995) against those of the NVSS survey (Condon et al., 1998), as can be seen in Figure \ref{fig:NVSS_FIRST}. 
The FIRST and NVSS surveys were carried out with the VLA in its B- and D-array
configurations respectively. Higher flux density values in D-array observations when
compared to those of B-array at the same frequency are an indicator
that there is undetected extended structure.

\begin{figure}
\nonumber
\centering
\includegraphics[width= \columnwidth]{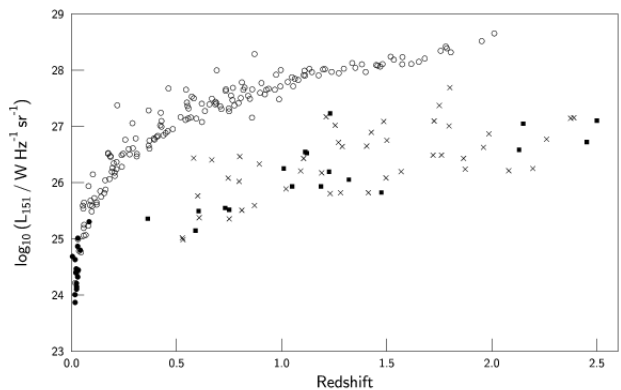}
\caption{\label{fig:3C7CPZ}Source luminosity at 151~MHz against redshift for the 3C survey (circles, with known FR I objects represented by the filled markers) and the 7C survey
(crosses, with the 18 quasars selected for this paper represented by the filled squares).} 
\end{figure}

In Figure \ref{fig:NVSS_FIRST} we see many objects with clear indications of extended structure. 
Given the finite amount of observing time available, we could neither observe all objects in the sample nor all objects with evidence of extended structure.
The sample we chose to observe is therefore in no sense complete but we believe it to be representative of the 7C quasar survey.
The 18 quasars we observed are listed in Table\,\ref{tab:obstable} which also lists their spectroscopically measured redshifts.\footnote{The sample of Riley et al. (1999) 
has almost complete optical spectroscopy providing confirmation (or in a few cases rejection) of the quasar
  hypothesis via detection of broad lines as well as measurements of redshifts. Spectra were gathered using, variously, the Isaac Newton Telescope (with the Faint Object Spectrograph, FOS-1), The William Herschel Telescope 
  (using ISIS and FOS-2) and the Nordic Optical Telescope (with Low-Dispersion Spectrograph).} Figure \ref{fig:3C7CPZ} shows the selected quasars in the $P$-$z$ plane. 

\begin{table*}
\centering
\begin{minipage}{110mm}
\caption{\label{tab:obstable}The sample of 18 quasars from Riley et al's (1999) quasar sample. This table
  shows the object names for both the J2000 and B1950 epochs as
  well as the central pointing coordinates used for the VLA
  observations in J2000 format, and the redshift of each quasar.
  The classification abbreviations are as follows: CD -- classical double, FD -- fat double, TJ -- FR I twin jet source, CJ -- core-jet source, WAT -- 
  wide angle tail (see Section \ref{sec:individual}), X -- undetermined .} 
\begin{tabular}{cccccc}\hline
\multicolumn{2}{c}{Source} & & & & \\
(J2000)           & (B1950)       & RA (J2000)   & Dec (J2000)& $z$   &Classification \\ \hline
1009+4655         & 1006+4710   & 10 09 40.601 & +46 55 23.03 & 1.010 &FD\\
1018+3542         & 1015+3557   & 10 18 10.989 & +35 42 39.27 & 1.226 &CJ\\
1021+4523         & 1018+4538   & 10 21 06.048 & +45 23 31.64 & 0.363 &FD\\
1022+3931         & 1019+3947   & 10 22 37.459 & +39 31 50.03 & 0.607 &CD\\
1023+3604         & 1020+3619   & 10 23 15.762 & +36 04 35.41 & 1.320 &TJ\\
1023+4414         & 1020+4429   & 10 23 29.805 & +44 14 14.21 & 0.750 &WAT?\\
1023+4824         & 1020+4839   & 10 23 29.794 & +48 24 36.99 & 1.231 &CD\\
1029+3224         & 1027+3239   & 10 29 59.132 & +32 24 19.42 & 1.187 &CJ\\
1030+3346         & 1027+3401   & 10 30 15.780 & +33 46 32.17 & 1.050 &TJ\\
1030+4309         & 1027+4324   & 10 30 21.547 & +43 09 07.84 & 1.112 &TJ\\
1032+4932         & 1029+4948   & 10 32 34.851 & +49 32 43.52 & 2.452 &CD\\
1033+4116         & 1030+4131   & 10 33 03.705 & +41 16 06.09 & 1.120 &X\\
1037+4650         & 1034+4705   & 10 37 15.391 & +46 50 14.31 & 2.150 &TJ\\
1038+3729         & 1035+3745   & 10 38 48.131 & +37 29 24.48 & 0.731 &FD\\
1040+4449         & 1037+4505   & 10 40 22.792 & +44 49 36.75 & 0.590 &TJ\\
1040+4529         & 1037+4545   & 10 40 42.316 & +45 29 39.57 & 1.475 &CJ\\
1054+4541         & 1051+4557   & 10 54 32.103 & +45 41 52.29 & 2.500 &TJ\\
1057+4556         & 1054+4612   & 10 57 18.453 & +45 56 15.16 & 2.130 &CD\\ \hline
\end{tabular}
\end{minipage}
\end{table*}

In an attempt to overcome the problems faced by previous radio
observations of quasars the data obtained for the sample
involved reasonably long integration times with all four configurations of the
VLA. We observed using the A-, B-, C- and D-arrays at 1.4~GHz and then
combined the data facilitating the sampling of the maximum possible
range of Fourier components, boosting the signal to noise
ratio of the final images and also favouring optimal
deconvolution.  

A summary of the observations
can be found in Table\,\ref{tab:vlatable}. The primary and
secondary calibrator sources were 1331+305 (3C\,286) and 1033+412
respectively. A single $\sim$2~minute observation of the primary
flux calibrator 1331+305 was carried out for each array
configuration and phase-calibrator 1033+412 was observed for $\sim$1~minute
intervals, approximately once per hour.

\begin{table}
\centering
\caption{\label{tab:vlatable}Summary of the VLA observations.}
\begin{tabular}{lll}\hline
Array         & Observation date & Approximate on-source \\ 
configuration &                  & integration time      \\ \hline 
A             & 2004 Sep 18      & 2~$\times$~10~minutes  \\
B             & 2003 Dec 19      & 2~$\times$~15~minutes  \\
C             & 2004 May 10      & 1~$\times$~15 minutes  \\
D             & 2004 Jul 29      & 2~$\times$~3~minutes   \\ \hline
\end{tabular}
\end{table}

Processing of the data was carried out using standard
procedures, including phase self-calibration, in {\sc aips}. The flux density scale was calibrated according to the methods of
Baars et al. (1977). Any suspicious or bad visibilities were
flagged using the {\sc tvflg} task in {\sc aips}. Combining and
cross-calibrating data from the different array
configurations using {\sc imagr}, {\sc calib} and {\sc dbcon} was
straightforward, although in certain cases strong confusing
sources had to be removed from the more compact configuration
data by means of subtraction of a clean component model ({\sc
imagr}, {\sc ccedt}, {\sc uvsub}). The final combined data sets
were cleaned and imaged using {\sc imagr}. Flux density
measurements were made using {\sc tvstat}, or {\sc imfit} in the
case of the unresolved sources given by certain compact array
observations.

\section{Results}
\label{sec:results}

\subsection{Radio maps}

% xfig font for radio maps = Helvetica Bold 17 point

\begin{figure*}
\begin{center}
\setlength{\unitlength}{1cm}
\begin{picture}(10,23.1)
\put(-4.0,-1.0){\includegraphics{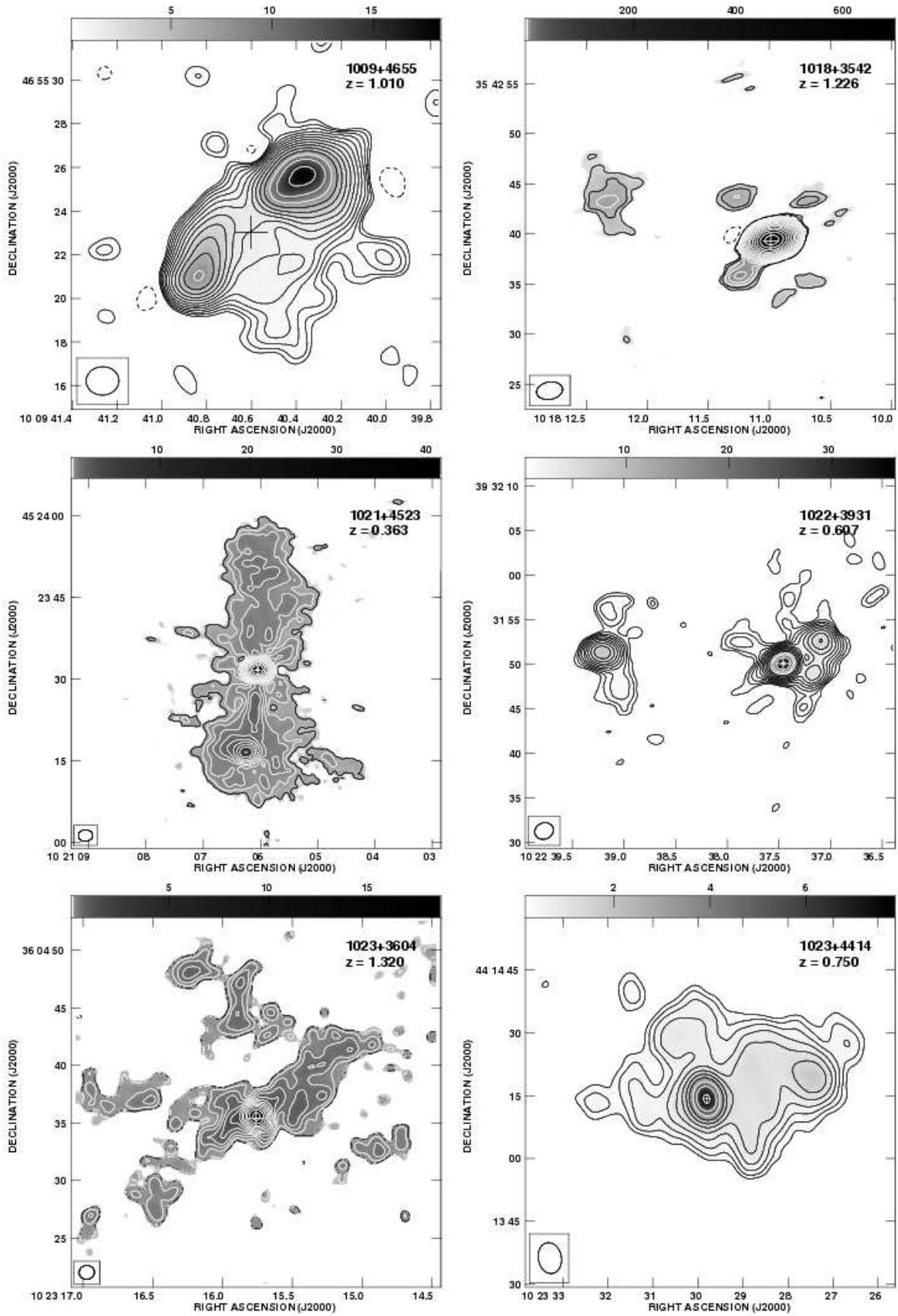}}
\end{picture}
\caption{\label{fig:radiomaps1}VLA images 1.}
\end{center}
\end{figure*}

\begin{figure*}
\nonumber
\begin{center}
\setlength{\unitlength}{1cm}
\begin{picture}(10,23.2)
\put(-3.0,0.5){\includegraphics{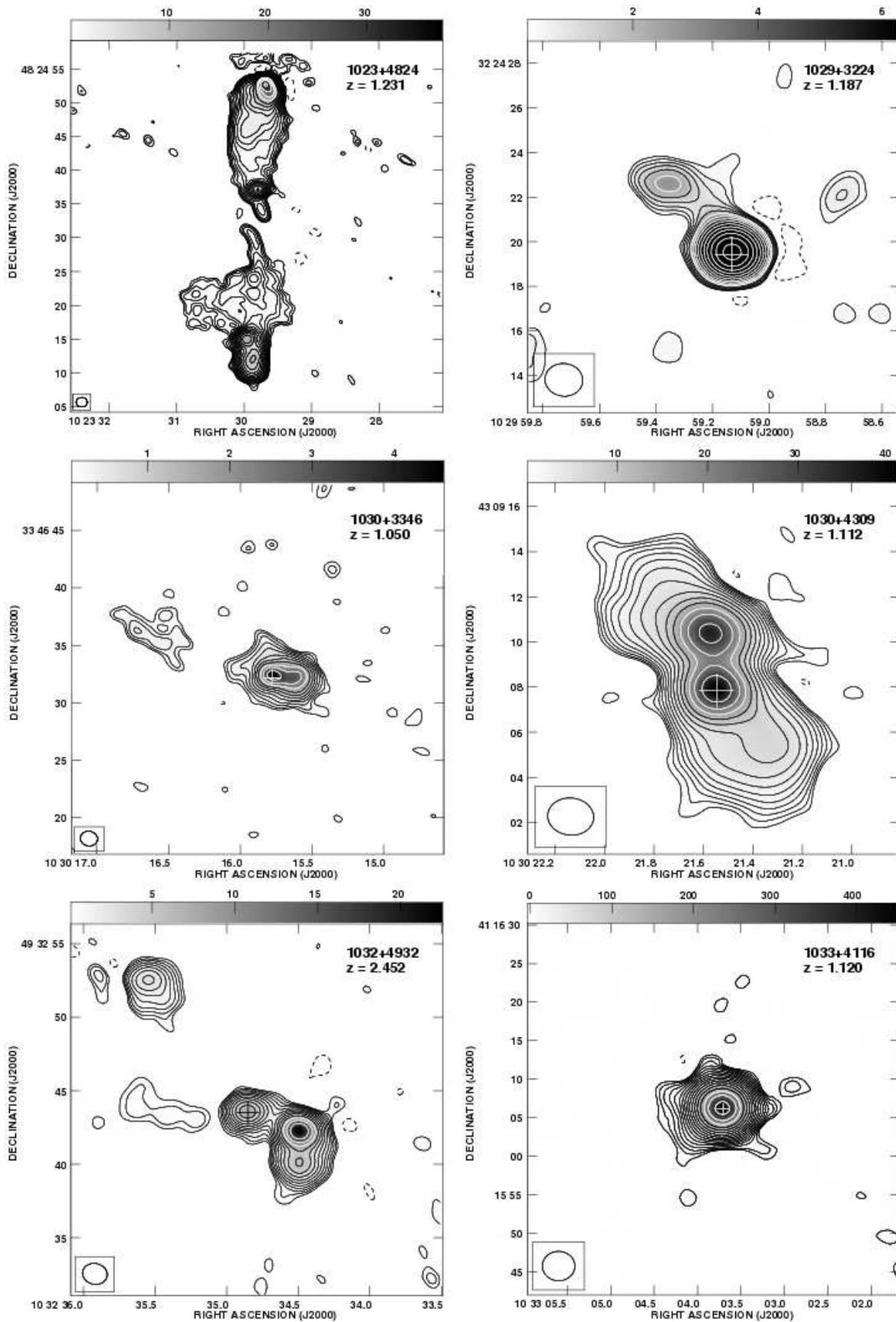}}
\end{picture}
\caption{\label{fig:radiomaps2}VLA images 2.}
\end{center}
\end{figure*}

\begin{figure*}
\nonumber
\begin{center}
\setlength{\unitlength}{1cm}
\begin{picture}(12,24.2)
\put(-3.0,0.5){\includegraphics{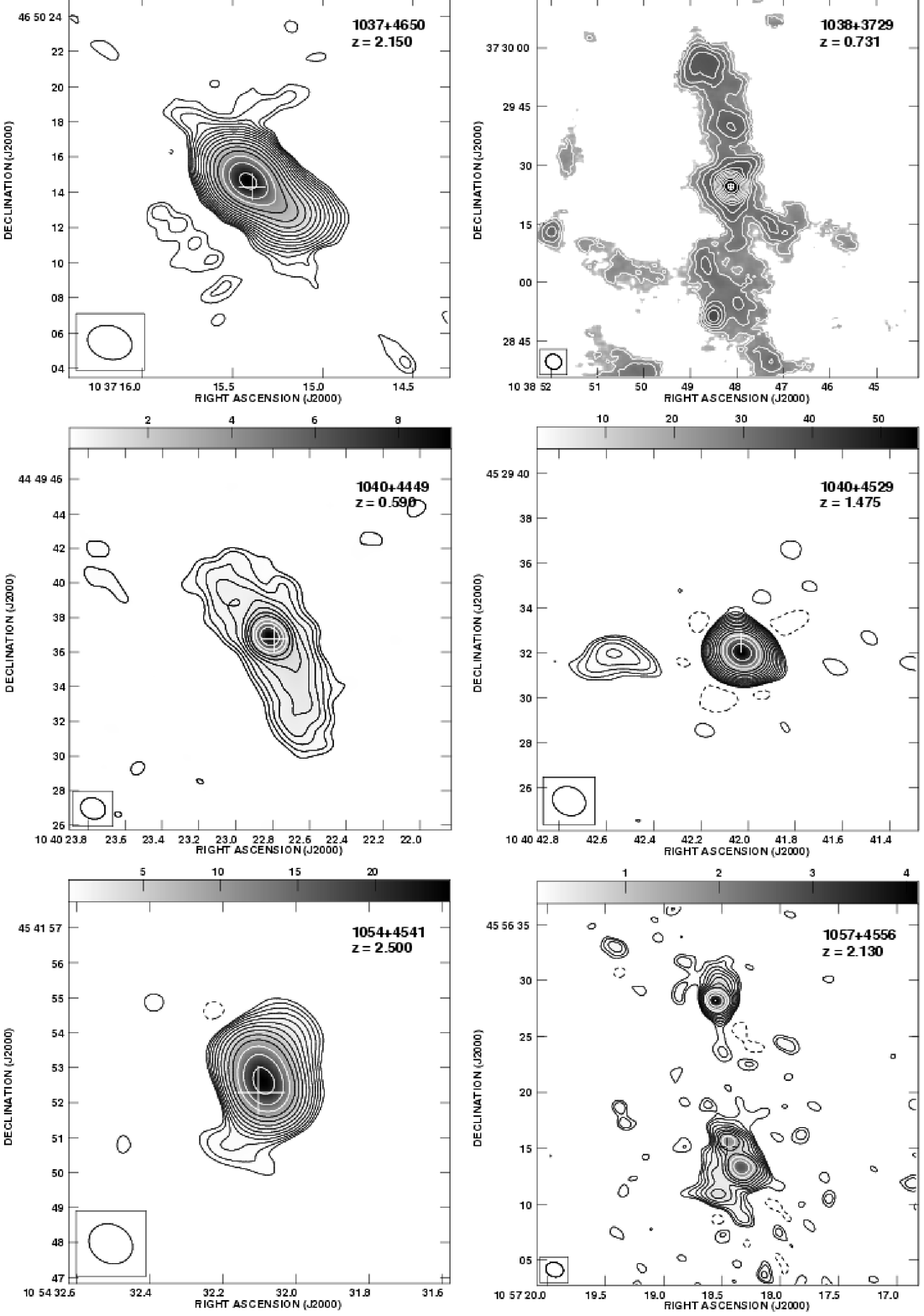}}
\end{picture}
%\vspace{465pt}
\caption{\label{fig:radiomaps3}VLA images 3.}
\end{center}
\end{figure*}

Figures\,\ref{fig:radiomaps1}, \ref{fig:radiomaps2} and \ref{fig:radiomaps3} show the combined
array 1.4 GHz radio images for the 18-quasar
sample. Table\,\ref{tab:mapstable} shows a summary of the final image
parameters. The fitted beam describes the half-power dimensions
and position angle of a 2D Gaussian fitted to the point spread
function for each image. In each case this is used as the
restoring beam when mapping. Each image is created with the maximum possible resolution (i.e.\ equivalent resolution of an A-array observation), except in the cases of 1023+4414 and 1038+3729 where the spatial resolution is degraded in order to highlight the interesting larger scale structures. This process is reflected in the fitted beam sizes listed in
Table\,\ref{tab:mapstable}.

For each image the lowest contour has a value of minus $\sqrt{2}$ times 2.5$\sigma$, where $\sigma$ is the rms background noise measured on each image. Following this single negative value, contours begin at 1 and increase in
multiples of $\sqrt{2}$ times ($\sigma$). The grey scale
pixel range runs from 2 times the rms noise to the peak flux density in
milli-Janskys per beam. The crosshairs on each map
show the central pointing coordinates as listed in
Table\,\ref{tab:obstable} and are coincident with the optical positions of the quasars. The crosshairs span 1.5 arcseconds.

\begin{table*}
\centering
\begin{minipage}{162mm}
\caption{\label{tab:mapstable}Summary of the radio maps presented
  in Figures\,\ref{fig:radiomaps1}, \ref{fig:radiomaps2} and
  \ref{fig:radiomaps3}. The contour
  multiplier is 2.5 times the rms background noise
  in each image and the grey scale pixel range runs from 2 times
  the rms noise to the peak flux in mJy / beam.}
\begin{tabular}{cccclll}\hline
Source       & Fitted beam        & PA        & Contour multiplier    & Greyscale range & Transfer  & Core visible in A-array\\ 
             & (arcseconds)       & (degrees) & mJy~/~beam            & mJy~/~beam        & function  & at 1.4~GHz? (Yes/No)\\ \hline
1009+4655    & 1.50~$\times$~1.30 & -87.12    & 0.175 & 0.14 -- 18.25 & Linear     &  No\\
1018+3542    & 2.60~$\times$~1.72 & -79.83    & 1.570 & 1.26 -- 694.95& Logarithmic&  Yes\\
1021+4523    & 2.63~$\times$~2.14 & -87.51    & 0.313 & 0.25 -- 41.21 & Logarithmic&  Yes\\
1022+3931    & 2.13~$\times$~1.81 & -68.39    & 0.389 & 0.31 -- 36.09 & Linear     &  Yes\\
1023+3604    & 1.34~$\times$~1.20 & -82.12    & 0.250 & 0.20 -- 18.49 & Logarithmic&  Yes\\
1023+4414    & 7.47~$\times$~5.50 & 10.72     & 0.231 & 0.19 -- 7.83  & Linear     &  Yes\\
1023+4824    & 1.62~$\times$~1.43 & 87.73     & 0.142 & 0.11 -- 37.62 & Linear     &  Yes\\
1029+3224    & 1.70~$\times$~1.46 & 83.31     & 0.424 & 0.34 -- 6.25  & Linear     &  Yes\\
1030+3346    & 1.50~$\times$~1.27 & 80.35     & 0.118 & 0.09 -- 4.59  & Linear     &  Yes\\
1030+4309    & 2.04~$\times$~1.65 & 85.07     & 0.350 & 0.28 -- 41.50 & Linear     &  Yes\\
1032+4932    & 1.69~$\times$~1.42 & 78.28     & 0.253 & 0.20 -- 22.47 & Linear     &  Yes\\
1033+4116    & 4.22~$\times$~3.82 & 89.33     & 1.538 & 1.23 -- 451.00& Linear     &  Yes\\
1037+4650    & 2.56~$\times$~1.90 & 74.74     & 0.276 & 0.22 -- 54.58 & Linear     &  Yes\\
1038+3729    & 4.15~$\times$~3.76 & 65.43     & 0.319 & 0.26 -- 16.29 & Logarithmic&  Yes\\
1040+4449    & 1.46~$\times$~1.21 & 70.04     & 0.207 & 0.14 -- 9.23  & Linear     &  Yes\\
1040+4529    & 1.51~$\times$~1.25 & 66.47     & 0.340 & 0.24 -- 55.23 & Linear     &  Yes\\
1054+4541    & 1.30~$\times$~1.07 & 61.00     & 0.338 & 0.27 -- 24.75 & Linear     &  Yes\\
1057+4556    & 1.60~$\times$~1.28 & 72.40     & 0.091 & 0.07 -- 4.11  & Linear     &  Yes\\ \hline
\end{tabular}
\end{minipage}
\end{table*}

\subsection{Radio flux density measurements}

Radio flux density measurements for the sample are
presented in Table\,\ref{tab:fluxtable} as determined by isolating the source structure in {\sc aips} with a box and executing the {\sc tvstat} verb. 
The error quoted is determined by the rms noise in the background of the image. The table contains the
1.4\,GHz flux densities as measured from the A-, B-, C- and
D-array observations. Note that the errors quoted are from map-specific flux density ratios which are accurate to better than one percent. Errors in the 
absolute flux values, due to the calibration according 
to the Baars scale, are typically a few percent (Perley, private communication and in prep.) and have not been included here.

The flux densities from the FIRST and NVSS
surveys are also presented for comparison. Generally for sources
with extended structure the flux-density measurements would be
expected to rise as the array becomes more compact and the
Fourier plane sampling becomes more favourable for the detection
of larger scale structures. This effect can be seen in several of
the sources from the sample (1021+4523, 1022+3931, 1023+4414,
1038+3729, 1040+4449, 1054+4541 -- see discussion in Section 4.1)
and in the cases where no compact hotspots are present this suggests they are of the FR\,I type. The NVSS survey was
conducted with the D-array and the flux-density values measured
from these observations are normally in good agreement with the D-array
measurements obtained for this study except in the case of variable sources, which we address in Section\,4.1 on a source by source basis. 
Similarly, except where noted in Table \ref{tab:fluxtable}, the B-array flux-density measurements
are in good agreement with the FIRST survey, also made using the VLA in B configuration.

\begin{table*}
\begin{center}
\begin{minipage}{144mm}
\caption{\label{tab:fluxtable}Flux density measurements for the
  sample of 18 7C quasars. Results from all four VLA array
  configurations are presented together with the values derived
  from the B-array FIRST survey and the D-array NVSS survey.} 
\begin{tabular}{@{\extracolsep{\fill}}lr@{$\pm$}lr@{$\pm$}lr@{$\pm$}lr@{$\pm$}llll}\hline
Source & \multicolumn{10}{c}{1.4\,GHz Flux Density (mJy)} & Notable flux differences \\
(J2000)  & \multicolumn{2}{c}{A-array} & \multicolumn{2}{c}{B-array} &
\multicolumn{2}{c}{C-array} & \multicolumn{2}{c}{D-array}             &FIRST& NVSS& for possible variability \\ \hline
1009+4655 & 71.5  & 0.2   & 69.0  & 0.9  & 74.7  & 0.2  & 70.0  & 0.6  & 70.6   & 77.3      & -- \\
1018+3542 & 676.6 & 3.9   & 650.0 & 3.0  & 695.0 & 1.0  & 682.0 & 3.4  & 594.1  & 615.2    & B/FIRST; D/NVSS \\
1021+4523 & 45.4  & 0.3   & 88.8  & 0.1  & 122.0 & 0.1  & 122.3 & 0.8  & 98.2   & 127.2    & -- \\
1022+3931 & 76.9  & 0.3   & 82.9  & 0.1  & 84.9  & 0.6  & 94.8  & 2.2  & 79.2   & 91.2    & -- \\
1023+3604 & 20.6  & 0.2   & 22.1  & 0.2  & 29.0  & 0.1  & 24.0  & 1.0  & 25.3   & 28.4      & -- \\
1023+4414 & 7.7   & 0.2   & 7.9   & 0.1  & 25.8  & 0.3  & 23.8  & 0.8  & 12.9   & 23.6      & -- \\
1023+4824 & 203.6 & 0.2   & 211.0 & 0.1  & 230.0 & 0.2  & 227.7 & 0.6  & 224.2  & 239.5   & -- \\
1029+3224 & 62.4  & 0.4   & 63.2  & 0.9  & 73.6  & 0.3  & 74.3  & 1.0  & 89.9   & 96.3     & B/FIRST; D/NVSS \\
1030+3346 & 14.5  & 0.1   & 14.3  & 0.1  & 18.5  & 0.2  & 16.7  & 1.2  & 14.1   & 16.4     & -- \\ 
1030+4309 & 124.0 & 0.2   & 120.7 & 0.4  & 126.6 & 0.2  & 124.4 & 1.3  & 125.9  & 129.9    & -- \\ 
1032+4932 & 47.1  & 0.2   & 39.3  & 0.1  & 45.6  & 0.1  & 42.0  & 0.9  & 44.3   & 46.8     & -- \\
1033+4116 & 436.8 & 2.4   & 458.8 & 0.7  & 518.2 & 0.2  & 515.8 & 1.6  & 432.7  & 473.2    & B/FIRST; D/NVSS \\
1037+4650 & 81.5  & 0.6   & 80.2  & 0.2  & 84.9  & 0.2  & 81.5  & 0.5  & 89.8   & 90.2   & B/FIRST; D/NVSS \\
1038+3729 & 14.7  & 0.3   & 31.2  & 0.3  & 43.8  & 0.5  & 55.0  & 1.2  & 29.2   & 54.3    & -- \\ 
1040+4449 & 9.2   & 0.3   & 15.4  & 0.1  & 23.0  & 0.1  & 20.8  & 1.0  & 14.7   & 23.0    & -- \\ 
1040+4529 & 53.6  & 0.5   & 50.9  & 0.2  & 55.2  & 0.1  & 52.7  & 0.8  & 51.8   & 53.4      & -- \\
1054+4541 & 44.6  & 0.4   & 47.3  & 0.2  & 52.2  & 0.2  & 51.5  & 1.0  & 49.7   & 52.3     & -- \\
1057+4556 & 9.1   & 0.1   & 9.3   & 0.1  & 17.3  & 0.1  & 15.9  & 0.2  & 15.8   & 14.9      & B/FIRST \\ \hline
\end{tabular} 
\end{minipage} 
\end{center}
\end{table*}

\section{Discussion}
\label{sec:discussion}

Whether a source is classified as an FR\,I or an FR\,II is determined by the morphology from
radio imaging (Fanaroff and Riley, 1974). If the source is edge brightened at the
extremities and with compact hotspots then it is generally of the
FR\,II (i.e.\ classical double) type. 

The radio maps presented in Figures\,\ref{fig:radiomaps1},
\ref{fig:radiomaps2} and \ref{fig:radiomaps3} show a variety of different
structures. Generally the objects can be classed as either
classical double FR\,II sources or the FR\,I type twin jet sources,
however there are also the more ambiguous structures which could
be classified as fat doubles (Owen \& Laing, 1989) or hybrid sources (Gopal-Krishna \& Wiita, 2002).  

A discussion of each quasar follows, including morphological
deductions from the radio images and the behaviour of the
measured flux density with $uv$ coverage. Certain objects have
also been previously observed at 8\,GHz using the VLA in
B-array. The features visible in these unpublished maps are
discussed where relevant. The increased resolution provided by
the high frequency observations is useful for identifying compact
regions of enhanced surface brightness. Features such as these
arising at the extremities of the object are thought to coincide
with the shock fronts occuring as the jets collide with the
ambient medium forming hotspots. These are generally taken as evidence that the kinetic energy
of the jet is still feeding the lobe head. Rounded, edge-brightened lobes which lack compact
hotspots are generally a property of a fat double source.
Objects are classified as FR\,I twin jet sources when they exhibit large-scale
extended radio structure that is similar to what is seen to arise from the twin-jet radio sources in the nearby Universe.

\subsection{Discussion of individual sources}
\label{sec:individual}

{\bf 1009+4655:} This source shows two bright lobes with
extended emission. There is no compact object visible at the
central pointing position. Examination of the 8-GHz image
reveals that the two peaks present in the 1.4-GHz image are each
resolved into elongated jet-like features as seen in some FR Is. These do not appear compact in the high resolution image. The axial ratio
means that this object would most likely be classified as a fat
double. The diffuse low brightness extensions to the south west
of the object may be plumes of emission diverted at right angles
to the jet axis.

{\bf 1018+3542:} The flux-density behaviour of this object points to high variability and does
not suggest any structure is being resolved out at any
stage. This source is by far the brightest object in the sample,
at almost 0.7\,Jy, and it is likely that the strength of the core
is limiting the dynamic range of the image making detection of
extended low surface brightness emission difficult. The features
which are detected hint at the presence of a highly asymmetric
structure, as is consistent with several other
objects in this sample (see below). This may be a consequence of
relativistic Doppler boosting.

{\bf 1021+4523:} The flux density values for this object
increase with decreasing array size suggesting undersampling of
smooth extended structure by the extended arrays. This would
ususally be an indicator of an FR\,I type object however the
southern part of this object appears to contain a compact hot
spot with a coherent structure linking it to the core. This
unusual object would probably be classified as a fat double, or
hybrid source. This object exhibits similarities with hybrid
sources such as 0131--367 (Gopal-Krishna \& Wiita, 2002) with
jets of FR\,I and FR\,II characteristics on different sides of the
nucleus. It is also noteworthy that the southern hotspot is recessed somewhat from the outer edge of the emission, as seen in some fat doubles (Owen \& Laing, 1989).

{\bf 1022+3931:} The central pointing coordinates are coincident
with the brightest of the four components visible in the map. The next brightest peak
to the north west of the core is most likely a hotspot and this object
is probably a FR\,II classical double. No emission linking the quasar to the peak visible to the east was found, even by varying the size of the restoring beam. It is likely however that this is emission from a second hotspot.

{\bf 1023+3604:} This source can be classified as a FR\,I twin
jet source even though the extended structure is very close to
the noise in the map. The extended emission on each side of the
core share a parallel axis even though they are slightly offset from the
core. This is characteristic of a single twist in the jets although precession of the jet axis may also be a possibility
(c.f.\ E\,1821+643, Blundell \& Rawlings, 2001, and 3C294, Erlund et al., 2006). 

{\bf 1023+4414:} The spatial resolution of this image was
deliberately degraded in order to reveal the extended
emission. Higher resolution imaging showed only the core and the
peak to the west, resolving out the diffuse envelope surrounding
them. The fifth positive contour hints at meandering emission with a comparatively large transverse angular size. While features such as these are indicative of wide angle tail sources, our classification for this object remains inconclusive.

{\bf 1023+4824:} The best example of a classical double FR\,II
structure in the sample, 1023+4824 shows extensive edge
brightening, a compact core and hotspots and asymmetrical
jets. The southern lobe shows what appears to be a region of
extended backflow. Classical doubles with large angular sizes will exhibit variations in flux density with larger baseline arrays, hence our selection criteria will not fully exclude objects such as this one. 

{\bf 1029+3224:} This source exhibits a variable core and the
flux density values hint at some extended structure being
resolved out. Maximum resolution imaging reveals a core-jet structure with a partially resolved
north eastern extension linked to the core.

{\bf 1030+3346:} Figure \ref{fig:radiomaps2} shows the maximum
resolution map for 1030+3346 showing FR\,I type emission coming
from the west of the core. The diffuse feature to the east of the
object is probably associated with the quasar. In a similar way
to the maximum resolution image of 1029+3224 the emission to the
east appears as a secondary low brightness peak connected to the
core in both the B-array data associated with this sample and the
VLA FIRST survey image.  This object is most likely to be a twin
jet FR\,I source, a claim backed up by comparison of the flux
density measurements (see Table \ref{tab:fluxtable}).

{\bf 1030+4309:} The
radio image of 1030+4309 shows extensive emission decreasing in
surface brightness away from the core, albeit on fairly small
(i.e.\ arcsecond) scales which explains the fairly steady flux measurements across different arrays.  The 8-GHz image reveals the core, the
bright peak 2.5 arcseconds to the north and a faint south western extension. The
radio morphology points to this source being an FR\,I quasar
although it is possible that the extended emission represents
plumes diverted from the hotspots at the ends of short jets.

{\bf 1032+4932:} All evidence suggests that this quasar has a distorted
classical double FR\,II structure associated with it. The flux
density remains steady for all array configurations. The 8-GHz
radio map shows the four components also visible in the 1.4-GHz
image although the southern extension and north eastern component
are very weak. The faint extended emission to the east may
represent diffuse emission associated with the quasar although
the radio emission is very much dominated by the core and the peak nearest to it.

{\bf 1033+4116:} Radio imaging of this object revealed only a
hint of an eastern extension at B-array, also seen in the FIRST
survey image. The cause of the variation in the flux density
values is most probably source variability. The true morphology of this source remains inconclusive.

{\bf 1037+4650:} The 1.4-GHz image shows a prominent south
western extension, and also an extension to the north east, suggesting this object is a FR\,I source. The
8-GHz image reflects this, showing the quasar and faint
extensions coincident with the ones visible at 1.4\,GHz and no evidence of any hotspots. 

{\bf 1038+3729:} Another probable fat double or hybrid source
with morphology remarkably similar to that of 1021+4523 showing a
pair of jets with a southern hotspot and a more diffuse northern
counter jet. The east-west feature seen running between the
core and the southern hotspot was coincident with the sidelobe
structure of a strong confusing source to the east although, after careful consideration, we think it is
unlikely that this feature is an imaging artifact due to its
curvature. It is uncertain whether it is associated with the
quasar, although radio galaxies have been seen to exhibit unusual
\emph{X}-shaped wing structures (e.g.~Dennett-Thorpe et al., 2002). The
stronger peak on the eastern edge of the map is coincident with a
nearby unrelated source. The jets of this source also appear to
be swept back slightly in an easterly direction, possibly a
result of motion through the ambient medium. 

{\bf 1040+4449:} This object has extended emission similar to
that of E\,1821+643 (Blundell \& Rawlings, 2001) and flux density measurements pointing to a
FR\,I classification. The slight bend in the major axis may
indicate some motion relative to the surrounding gaseous medium in a north-westerly direction.

{\bf 1040+4529:} It is not possible to draw a definite conclusion as
to the nature of this object. The feature to the east of the
quasar may be a faint jet and the VLA FIRST survey image hints at
a eastern extension, however the flux density values suggest that
if there is any diffuse extended emission it is below
detectability for the observations presented in this paper.

{\bf 1054+4541:} At a redshift of 2.5 this is the most distant
object in the sample. Flux density measurements from different VLA arrays suggest there is extended structure associated with this
object and there is a hint of oppositely directed jet activity in the radio image. The extended features in this image are real, but are not
coincident with any compact features in a high-resolution MERLIN image (Heywood et al., in prep.).

{\bf 1057+4556:} The final object in the sample is another
highly asymmetric FR\,II classical double. It is in many was
morphologically similar to 1022+3931 and 1032+4932, with large
differences in separation between the core and each hotspot and
what appears to be a diverted plume associated with one hotspot.  

\subsection{Radio luminosity and redshift considerations}

\begin{figure}
\nonumber
\centering
\includegraphics[width= \columnwidth]{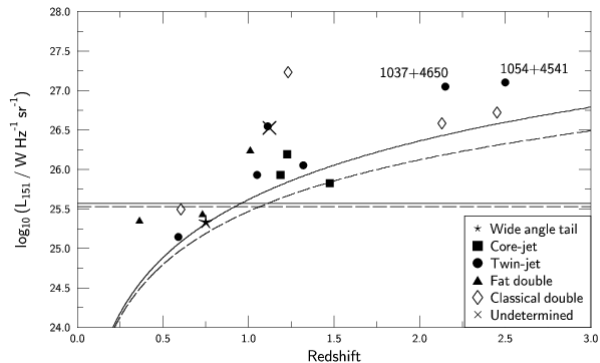}
\caption{\label{fig:rad_z}The 151-MHz radio luminosity against
  redshift. The horizontal lines show the traditional FR\,I / FR
  II break luminosity adjusted for a frequency of 151~MHz, and
  the curved lines show the flux density limit as a function of
  redshift for a 0.1-Jy source. For both of these indicators the
  assumed spectral indices are 0.5 and 1.0 for the dashed and
  solid lines respectively.} 
\end{figure}

Figure\,\ref{fig:rad_z} shows 151-MHz radio luminosity
plotted against redshift for the objects in the
sample. The horizontal lines show the traditional FR\,I / FR\,II
break luminosity at 151~MHz assuming spectral indices of 0.5
(dashed) and 1.0 (solid). The curved lines show the detection
limits as a function of redshift for a source of flux density
0.1~Jy for assumed spectral indices of 0.5 and 1.0.   

By examining this plot it is evident that there exist FR\,I type
sources over a large range in luminosity and above the traditional FR\,I /
FR\,II break. If 1037+4650 and 1054+4541 (the labelled, filled circles on Figure \ref{fig:rad_z}) are indeed bona fide FR\,I
sources then the luminosity range for FR\,I type quasars extends
to over 10$^{27}$~W~Hz$^{-1}$~sr$^{-1}$ in the cosmology assumed in this paper. Whether these
objects are FR\,I sources is not completely certain however it is
noteworthy that they are definitely do not have compact hotspots associated with edge-brightened extremities, as would be
characteristic of a classical double source.

There is the possibility that these high redshift objects
could resemble, for example, B2 1108+27 (Owen et al., 2000). This
source initially appeared to be an arcminute-scale FR\,I-type
structure on the basis of the NVSS image, however a two hour integration with the VLA in D-array
revealed that the emission associated with the jets of this
object actually extended to an angular size of 30 arcminutes,
corresponding to a 1-Mpc structure. It is possible that the
radio images of sources such as 1037+4650 and 1054+4541 presented
in this paper simply represent the inner regions of sources similiar
to B2\,1108+27. Detection of faint, extended emission around these sources could only be confirmed with
subsequent very deep, pointed compact-array observations.

Another consideration is the fact that quasars have bright cores and possibly exhibit very strongly beamed emission. Strongly beamed jets may influence the classification
of an object by concentrating the emission into a small angular size. Our initial selection however was not made at a frequency at which Doppler boosting operates (e.g. 1.4~GHz), but 
at the lower frequency of 151~MHz which measures on isotropic lobe emission. 

If a more cautious approach is adopted and it is assumed that
1037+4650 and 1054+4541 cannot be considered to be FR\,I sources,
there are still FR\,I-type twin-jet sources with luminosities of
up to 10$^{26.4}$~W~Hz$^{-1}$~sr$^{-1}$ (1030+4309), almost a factor
of 10 above the traditional FR\,I / FR\,II break. 

Furthermore, although FR\,I type sources are observed with
luminosities over a factor of ten higher than the traditional
FR\,I / FR\,II break, there are no FR\,II objects (with the
borderline exception of 1022+3931) with luminosities
significantly below the break.  For this reason we believe that
the FR\,I / FR\,II break is still significant as a threshold
below which classical doubles are not in general found, although
clearly a deeper understanding of FR I evolution is required,
building on the existing work on low-luminosity radio sources
(e.g.\ Clewley \& Jarvis, 2004).

A second possible explanation for the presence of these 
high-luminosity FR\,I quasars arises by considering jet power and the
environment surrounding the object. A low power jet will only
give rise to an FR\,I type structure since the flow is
subsonic. Shock fronts do not occur when the material collides
with the intergalactic medium and therefore no hotspots are ever
formed at the jet extremities. It is postulated here that high
power jets can however give rise to both FR\,I and FR\,II
morphologies depending on their environment. For a smooth and
rarefied local environment a high power jet can continue
relatively unhindered and remain collimated until it strikes the
boundary with the IGM, resulting in shock fronts and the
increased brightness hotspots which characterise a FR\,II classical double. For a
dense or clumpy local environment the jet may lose much of its
momentum and dissipate its energy in a more gradual fashion
resulting in an FR\,I type object.

At higher redshift the enivronments surrounding active galactic
nuclei may well be denser and clumpier than those of the local
Universe. Jets associated with high redshift environments would
be more prone to disruption than those associated with objects in
the local Universe, thus only well collimated, high power jets
would produce FR\,II type structures (Barthel \& Miley, 1988).  

\section{Conclusions}
\label{sec:conclusions}

The most obvious conclusion that can be drawn from these results
is that E1821+643 is not a unique exception as an FR\,I quasar
and that FR\,I quasars not uncommon amongst populations selected to be both radio-loud and quasars: objects with broad emission lines can give rise to FR\,I
structures, which apparently have low power radio jets.

The two classes of radio morphology described by Fanaroff \&
Riley (1974) appear to apply to quasars, in that quasars exist in
both FR\,I and II categories. However, our results also show
that there is no sharp transition across a certain luminosity as
we observe FR\,I quasars with luminosities at least a factor of 10
above the traditional FR\,I / FR\,II break.

The source 1023+4824 is the only
object in the sample to show a \emph{clear} classical double
morphology. Three sources which are dominated by core and
hotspot emission and could therefore be thought of as hybrid FR\,II
classical double sources exhibit obvious similarities (1022+3931,
1032+4932 and 1057+4556). All three of these feature a double
hotspot and are highly asymmetric about the core. Generally the
sources are either FR\,I twin jet (or FR\,I-\emph{like}) sources
(e.g.\ 1023+3604, 1030+3346, 1030+4309, 1037+4650, 1040+4449) or unusual hybrid or fat double sources
(e.g.\ 1009+4655, 1021+4523, 1038+3729).  

Most objects observed at $z\sim0.6$ in the 7C quasar survey are FR\,Is (see Figure \ref{fig:rad_z}),
and lie below the traditional FR\,I/FR\,II break in luminosity. The lack of FR\,I quasars in
the 3C survey may be due, at least in part, to rarity issues. Although the comoving volumes probed by 3CRR and 7C 
are comparable (3CRR probes $\sim$1~$\times$~10$^{8}$~Mpc$^{3}$ to a redshift of 0.1, 7C probes $\sim$2~$\times$~10$^{8}$~Mpc$^{3}$ 
to a redshift of 0.6), quasars are more numerous at $z\sim0.6$ than at $z\sim0.1$.
However, it does seem from studies of complete samples of radio sources at $z\sim0.6$
(McLure et al., 2004; Mitchell, 2005) that a typical L$_{151}\sim$10$^{(25-25.5)}$~W~Hz$^{-1}$~sr$^{-1}$ radio source at $z\sim0.6$ has a FR\,I radio structure but 
has neither an observed quasar nucleus (as signified by, for example, the presence of broad H$\alpha$ emission lines) nor evidence for
any buried nucleus of comparable intrinsic luminosity (inferred, e.g., from strong narrow emission lines). 

Again with reference to Figure \ref{fig:rad_z}, at $z\sim1.25$ we seem to generally observe FR I-like structures.
In the 7C quasar survey, FR I objects exist at higher radio luminosities (10$^{(26-26.5)}$~W~Hz$^{-1}$~sr$^{-1}$) than those
observed in the 3C survey at low redshift. 
There is thus evidence that the division between FR\,I and FR\,II radio sources changes
with $z$, in the sense that FR\,Is can exist above the traditional break luminosity at higher redshifts.
The obvious explanation for this (which is therefore also likely to affect the $z\sim0.5$ population) is
a systematic change in the environments of radio sources with cosmic epoch (see Section 4.2).

For the objects at $z\sim2.3$ the data presented here have shown that the furthest extents of these sources are not coincident with hotspots, but it
remains open as to whether these objects are truly FR Is or have, for instance, disrupted or precessing jets. Further high resolution radio imaging would be useful
to resolve such questions.

Disproving the idea that FR\,I radio structures cannot be associated with a quasar (i.e. an acccreting
black hole central engine) is quite an important conclusion. It has recently become clear from surveys with Spitzer (e.g. Martinez-Sansigre et al., 2005)
that most quasar activity, and hence black hole growth, in the universe has been missed by optical and X-ray surveys due to obscuration by dust and gas.
Many of these obscured quasars have radio luminosities and spectral indices consistent with the hypothesis that they are associated
with FR\,I-like radio jets (Martinez-Sansigre et al., 2006). The objects observed in this paper may be unobscured examples of the same phenomenon. The growth of supermassive black holes may thus go hand-in-hand with FR\,I-like jet
output which may have critical effects on the evolution of galaxies (e.g. Croton et al., 2005) and larger-scale structures.

\section*{Acknowledgements}

IH and SR thank PPARC and KMB thanks the Royal Society for a
University Research Fellowship. We also thank the referee for useful comments. The VLA is a facility of the
NRAO operated by Associated Universities, Inc., under
co-operative agreement with the National Science Foundation. This research has
made use of NASA's Astrophysics Data System.

\bsp % ``This paper has been produced using the ...''

\label{lastpage}


\begin{thebibliography}{}

\bibitem[Baars et al.(1977)]{1977A&A....61...99B} Baars, J.~W.~M., Genzel, 
R., Pauliny-Toth, I.~I.~K., \& Witzel, A.\ 1977, \aap, 61, 99 

\bibitem[Barthel(1989)]{1989ApJ...336..606B} Barthel, P.~D.\ 1989, \apj, 
336, 606

\bibitem[Barthel \& Miley(1988)]{1988Natur.333..319B} Barthel, P.~D., \& 
Miley, G.~K.\ 1988, \nat, 333, 319

\bibitem[Baum et al.(1995)]{1995ApJ...451...88B} Baum, S.~A., Zirbel, 
E.~L., \& O'Dea, C.~P.\ 1995, \apj, 451, 88 

\bibitem[Becker et al.(1995)]{1995ApJ...450..559B} Becker, R.~H., White, 
R.~L., \& Helfand, D.~J.\ 1995, \apj, 450, 559 

\bibitem[Blundell(2005)]{Blundell2005} Blundell, K.~M., 2005, Phil.\
  Trans.\ Roy.\ Soc., 363, 645

\bibitem[Blundell \& Rawlings(2001)]{2001ApJ...562L...5B} Blundell, K.~M., 
\& Rawlings, S.\ 2001, \apjl, 562, L5 

\bibitem[Blundell 
.(1999)]{1999AJ....117..677B} Blundell, K.~M., 
Rawlings, S., \& Willott, C.~J.\ 1999, \aj, 117, 677

\bibitem[Clewley \& Jarvis(2004)]{2004MNRAS.352..909C} Clewley, L., \& 
Jarvis, M.~J.\ 2004, \mnras, 352, 909 

\bibitem[Condon et al.(1998)]{1998AJ....115.1693C} Condon, J.~J., Cotton, 
W.~D., Greisen, E.~W., Yin, Q.~F., Perley, R.~A., Taylor, G.~B., \& 
Broderick, J.~J.\ 1998, \aj, 115, 1693

\bibitem[Croton et al.(2005)]{2005MNRAS.356.1155C} Croton, D.~J., et al.\ 
2005, \mnras, 356, 1155 

\bibitem[Dennett-Thorpe 
.(2002)]{2002MNRAS.330..609D} Dennett-Thorpe, 
J., Scheuer, P.~A.~G., Laing, R.~A., Bridle, A.~H., Pooley, G.~G., \& 
Reich, W.\ 2002, \mnras, 330, 609 

\bibitem[Erlund et al.(2006)]{2006MNRAS.371...29E} Erlund, M.~C., Fabian, 
A.~C., Blundell, K.~M., Celotti, A., \& Crawford, C.~S.\ 2006, \mnras, 371, 
29 

\bibitem[Falcke et al.(1995)]{1995A&A...298..395F} Falcke, H., 
Gopal-Krishna, \& Biermann, P.~L.\ 1995, \aap, 298, 395

\bibitem[Fanaroff \& Riley(1974)]{1974MNRAS.167P..31F} Fanaroff, B.~L., \& 
Riley, J.~M.\ 1974, \mnras, 167, 31P 

\bibitem[Goldschmidt et al.(1999)]{1999ApJ...511..612G} Goldschmidt, P., 
Kukula, M.~J., Miller, L., \& Dunlop, J.~S.\ 1999, \apj, 511, 612

\bibitem[Gopal-Krishna \& Wiita(2002)]{2002NewAR..46..357G} Gopal-Krishna, 
\& Wiita, P.~J.\ 2002, New Astronomy Review, 46, 357 

\bibitem[Kaneda et al.(1995)]{1995ApJ...453L..13K} Kaneda, H., et al.\ 
1995, \apjl, 453, L13

\bibitem[Hooper 
.(1996)]{1996ApJ...473..746H} Hooper, E.~J., Impey, 
C.~D., Foltz, C.~B., \& Hewett, P.~C.\ 1996, \apj, 473, 746 

\bibitem[Kellermann et al.(1994)]{1994AJ....108.1163K} Kellermann, K.~I., 
Sramek, R.~A., Schmidt, M., Green, R.~F., \& Shaffer, D.~B.\ 1994, \aj, 
108, 1163 

\bibitem[Kukula et al.(1998)]{1998MNRAS.297..366K} Kukula, M.~J., Dunlop, 
J.~S., Hughes, D.~H., \& Rawlings, S.\ 1998, \mnras, 297, 366 

\bibitem[Mart{\'{\i}}nez-Sansigre et al.(2005)]{2005Natur.436..666M} 
Mart{\'{\i}}nez-Sansigre, A., Rawlings, S., Lacy, M., Fadda, D., Marleau, 
F.~R., Simpson, C., Willott, C.~J., \& Jarvis, M.~J.\ 2005, \nat, 436, 666 

\bibitem[Mart{\'{\i}}nez-Sansigre et al.(2006)]{2006MNRAS.373L..80M} 
Mart{\'{\i}}nez-Sansigre, A., Rawlings, S., Garn, T., Green, D.~A., 
Alexander, P., Kl{\"o}ckner, H.-R., \& Riley, J.~M.\ 2006, \mnras, 373, L80 

\bibitem[McLure et al.(2004)]{2004MNRAS.351..347M} McLure, R.~J., Willott, 
C.~J., Jarvis, M.~J., Rawlings, S., Hill, G.~J., Mitchell, E., Dunlop, 
J.~S., \& Wold, M.\ 2004, \mnras, 351, 347 

\bibitem[Miller et al.(1990)]{1990MNRAS.244..207M} Miller, L., Peacock, 
J.~A., \& Mead, A.~R.~G.\ 1990, \mnras, 244, 207 

\bibitem[Miller et al.(1993)]{1993MNRAS.263..425M} Miller, P., Rawlings, 
S., \& Saunders, R.\ 1993, \mnras, 263, 425 
 
\bibitem[Mitchell(2005)]{Mitchell2005}Mitchell, E.~K.,\ 2005, PhD thesis, Univ. Oxford

\bibitem[Owen \& Laing(1989)]{1989MNRAS.238..357O} Owen, F.~N., \& Laing, 
R.~A.\ 1989, \mnras, 238, 357 

\bibitem[Owen et al.(2002)]{2002IAUS..199..171O} Owen, F.~N., Ledlow, 
M.~J., Eilek, J.~A., Kassim, N.~E., Miller, N., Dwarakanath, K.~S., \& 
Ivison, R.~J.\ 2002, The Universe at Low Radio Frequencies, 199, 171 

% \bibitem[Owen et al.(2000)]{2000astro.ph..6152O} Owen, F.~N., Ledlow, 
% M.~J., Eilek, J.~A., Kassim, N.~E., Miller, N.~A., Dwarakanath, K.~S., \& 
% Ivison, R.~J.\ 2000, ArXiv Astrophysics e-prints, arXiv:astro-ph/0006152

\bibitem[Pearson et al.(1992)]{1992MNRAS.259P..13P} Pearson, T.~J., 
Blundell, K.~M., Riley, J.~M., \& Warner, P.~J.\ 1992, \mnras, 259, 13P 

\bibitem[Rawlings \& et al.(1998)]{1998ocnr.conf..171R} Rawlings, S., \& et 
al.\ 1998, ASSL Vol.~226: Observational Cosmology with the New Radio 
Surveys, 171 

\bibitem[Riley et al.(1999)]{1999MNRAS.307..293R} Riley, J.~M., Rawlings, 
S., McMahon, R.~G., Blundell, K.~M., Miller, P., Lacy, M., \& Waldram, 
E.~M.\ 1999, \mnras, 307, 293

\bibitem[White et al.(2000)]{2000ApJS..126..133W} White, R.~L., et al.\ 
2000, \apjs, 126, 133

\end{thebibliography}
\end{document}